\documentclass[twocolumn, secnumarabic, amssymb, nobibnotes, aps]{revtex4-1}
\usepackage{amsmath,amstext,bm,appendix,graphicx,multirow,amsfonts}
\usepackage{color}
\usepackage{diagbox}
\usepackage[bookmarks=false]{hyperref}
\hypersetup{colorlinks=true,citecolor=blue,
linkcolor=blue,urlcolor=blue,pdfstartview=FitH,bookmarksopen=true}

\begin{document}
\title{Resonant-interaction-induced Rydberg antiblockade}
\author{J. L. Wu}
\affiliation{School of Physics and Engineering, Zhengzhou University, Zhengzhou 450001, China}
\affiliation{Department of Physics, Harbin Institute of Technology, Harbin 150001, China}
\author{S. L. Su}
\email[]{slsu@zzu.edu.cn}
\affiliation{School of Physics and Engineering, Zhengzhou University, Zhengzhou 450001, China}
\author{J. Song}
\email[]{jsong@hit.edu.cn}
\affiliation{Department of Physics, Harbin Institute of Technology, Harbin 150001, China}

\begin{abstract}
Different from the conventional Rydberg antiblockade~(RAB) regime that either requires weak Rydberg-Rydberg interaction~(RRI), or compensates Rydberg-Rydberg interaction~(RRI)-induced energy shift by introducing dispersive interactions, we show that RAB regime can be achieved by resonantly driving the transitions between ground state and Rydberg state under strong RRI. The Rabi frequencies are of small amplitude and time-dependent harmonic oscillation, which plays a critical role for the presented RAB. The proposed unconventional RAB regime is used to construct high-fidelity controlled-Z~(CZ) gate and controlled-not~(CNOT) gate in one step. Each atom requires single external driving. And the atomic addressability is not required for the presented unconventional RAB, which would simplify experimental complexity and reduce resource consumption.
\end{abstract}
\maketitle

\section{Introduction}
Rydberg atom denotes the neutral atom that could be excited to high-lying Rydberg states, which would exhibit large dipole moments when they are excited and close enough~\cite{Gallagher1994,PhysRevLett.85.2208,PhysRevLett.87.037901} and have been well studied in quantum information science~\cite{Saffman2010,saffman2016quantum}. The Rydberg-Rydberg interaction~(RRI) between excited Rydberg atoms has been directly measured in experiments~\cite{PhysRevLett.110.263201}. The traditional viewpoint is that a frequency-matched~(resonant) classical laser cannot excite more than one Rydberg atoms at the same time~\cite{PhysRevLett.85.2208,PhysRevLett.87.037901}(known as Rydberg blockade) when the distance among them is less than the blockade radius and the line width of the excitation is significantly narrower than the energy shift caused by the RRI. Rydberg blockade between two atoms located about 10 $\mu$m~\cite{urban2009observation} apart
through sequential driving and 4 $\mu$m~\cite{gaetan2009observation} apart by collective
driving has been observed in experiments, respectively.

In contrast to Rydberg blockade, Rydberg antiblockade~(RAB) allows more than one Rydberg atom to be excited~\cite{Ates2007,Pohl2009,qian2009,Amthor2010} and has also been well studied~\cite{PhysRevLett.85.2208,Zuo2010,Lee2012,Carr2013,Li2013,Su2016,*Su201702,Su2018,Shao2017,chen2018accelerated,Li:18,Li2019} for Rydberg-atom-based preparation of quantum entangled state and construction of quantum logic gate. Conventionally, RAB can be classified as two types. The first type works when the value of the strong Rabi frequency $\Omega$ is much higher than that of the weak RRI strength \emph{V}~\cite{PhysRevLett.85.2208}, based on which the two-qubit Rydberg quantum logic gate can be constructed in three steps~\cite{PhysRevLett.85.2208}. Step one is to excite the state $|11\rangle$ to two-excitation Rydberg state $|rr\rangle$; Step two is to wait for time duration $\tau$, and $|rr\rangle$ would get a phase $\theta=Vt$. The third step is the inverse step of the first step. The other states $|00\rangle$, $|01\rangle$, and $|10\rangle$ keep invariant because the RRI is inexistent during the three steps. The second type is to use the blue-detuned laser to compensate the RRI with the condition $\sum_{i}\Delta_{i}=\sum_{j}V_{j}$~\cite{Zuo2010,Lee2012,Carr2013,Li2013,Su2016,*Su201702,Su2018,Shao2017,chen2018accelerated,Li:18,Li2019}, where $\Delta_{i}$ denotes the detuning for the \emph{i}-th atom and $V_{j}$ denotes the RRI strength for the \emph{j}-th pair of Rydberg atoms. After the pioneering works in Refs.~\cite{Ates2007,Pohl2009,qian2009,Amthor2010}, the strict condition to achieve the Rabi oscillation between two-atom ground state and the two-excitation Rydberg state was studied~\cite{Zuo2010,Lee2012,Carr2013,Li2013}. And the RAB was used to prepare entangled and antiferromagnetic states by dissipative dynamics~\cite{Carr2013}. Also, the dynamics of the direct laser excitation of a pair of
Rydberg atoms was also studied based on this type of RAB~\cite{Li2013}. Then this type of RAB was generalized to multiple-qubit case under asymmetry RRI~\cite{Su2018} and used to construct quantum logic gate~\cite{Su2016,Su201702,Su2018} and prepare quantum entangled state~\cite{Shao2017,Li2019}.

Different from the conventional RAB regimes mentioned above, in this letter we show that, through modulating the Rabi frequency, a frequency-matched resonant classical laser can excite more than one Rydberg atoms under strong RRI in one step. Then the proposed unconventional RAB regime is used to construct the controlled-Z~(CZ) and controlled-NOT~(CNOT)~gates, respectively.
The present scheme has the following properties. (i) The proposed RAB regime is induced by resonant and small-amplitude driving. It is different from the two types of conventional RAB mentioned in the above paragraph. (ii) Based on the proposed RAB, the high fidelity CZ gate and CNOT gate can be constructed in one step, which may enhance the efficiency of Rydberg-atom-based quantum computation. (iii) The proposed RAB and CZ gates are of less resource consumption because the scheme requires only one external driving and does not require atomic addressability.

\begin{figure}
\includegraphics[width=\linewidth]{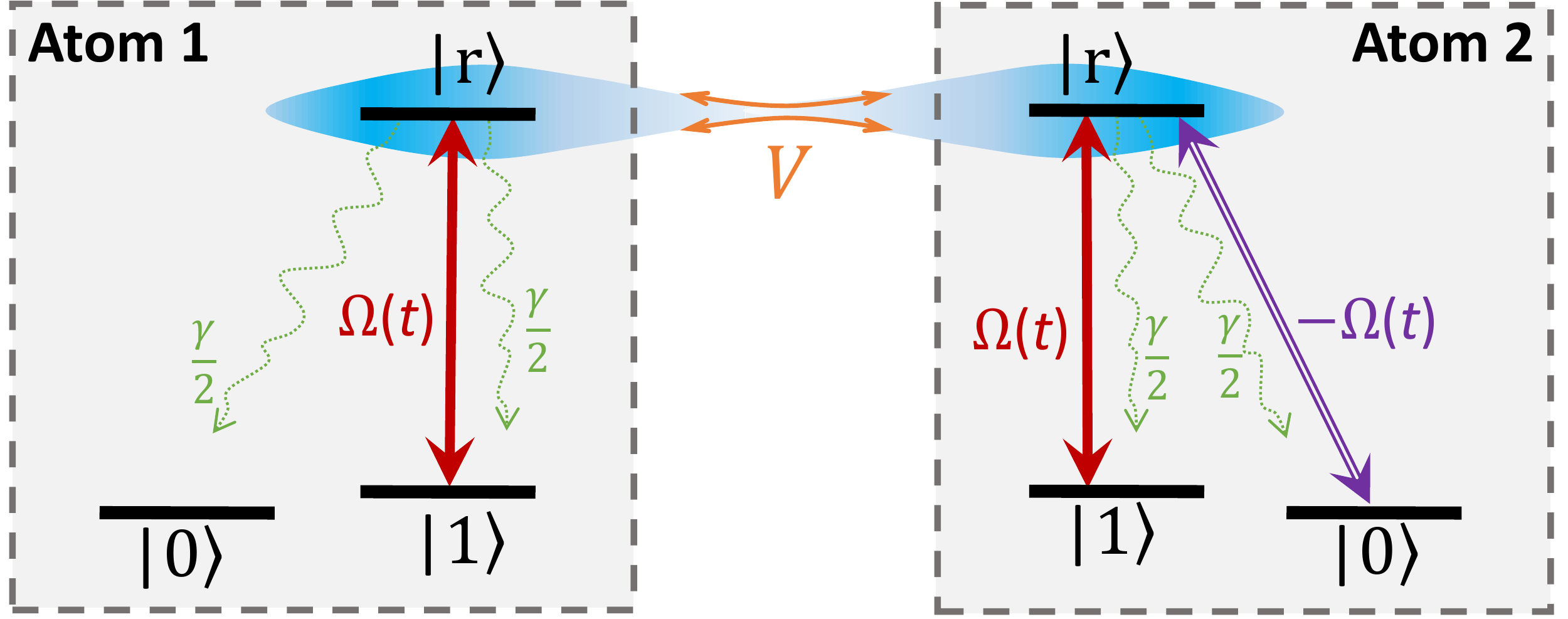}\\
\caption{The diagrammatic sketch of the Rydberg-atom system for realizing the unconventional RAB and CZ gate. Identical level structure in two atoms is used, including one Rydberge state $|r\rangle$ and two steady lower-energy states $|0\rangle$ and $|1\rangle$. External classical fields are imposed to resonantly drive the atomic transition $|1\rangle\leftrightarrow|r\rangle$ with the time-dependent Rabi frequency $\Omega(t)$. $V$ denotes the RRI strength. Another classical field imposed on Atom~2 and resonantly driving the transition $|0\rangle_2\leftrightarrow|r\rangle_2$ with the Rabi frequency $-\Omega(t)$ is introduced for constructing the CNOT gate. Here we assume that the spontaneous emission rates from $|r\rangle$ to $|0\rangle$ or $|1\rangle$ for the two atoms are $\gamma/2$ equivalently.}\label{F1}
\end{figure}
The considered configuration of the scheme was shown in Fig.~\ref{F1}, where two Rydberg atoms are trapped by means of optical tweezers. Each atom has one Rydberg state $|r\rangle$ with high principal quantum number and two ground states $|0\rangle$ and $|1\rangle$. The coupling between ground state $|1\rangle$ and Rydberg state $|r\rangle$ can be achieved via two ways. One way is the two-photon process, which has been demonstrated in Refs.~\cite{urban2009observation,gaetan2009observation,PhysRevLett.104.010503,PhysRevA.82.030306}. In this way the atomic energy levels are chosen as $|0\rangle\equiv|5s_{1/2}, f = 1, m_{f}=0\rangle$, $|1\rangle\equiv|5s_{1/2}, f = 2, m_{f}=0\rangle$, $|r\rangle\equiv|97d_{5/2}, m_{j}=5/2\rangle$, and the intermediate state is chosen as $|p\rangle = |5p_{3/2},f'=2\rangle$ of $^{87}$Rb. Meanwhile, for the two-photon process, the coherent excitation from ground state to Rydberg state with an adjustable effective Rabi frequency has also been demonstrated very recently~\cite{PhysRevA.97.053803} through tuning of the coupling strength between the ground state and the intermediate state. The other way is the single-photon process, which has been demonstrated in Ref.~\cite{jau2016entangling} by considering $^{133}$Cs atoms. In that way, the high-lying Rydberg state $|r\rangle\equiv|64P_{3/2};m_J=3/2\rangle$ couples directly to the ground state by a single exciting laser. The two ground states are magnetically insensitive ``clock" states $|0\rangle\equiv|6S_{1/2};F=3,M_F=0\rangle$ and $|1\rangle\equiv|6S_{1/2};F=4,M_F=0\rangle$~\cite{jau2016entangling}.

In our scheme, the two Rydberg atoms interact with each other through the RRI with the strength $V$. Here we do not consider the transition $|0\rangle_2\rightarrow|r\rangle_2$ of the Atom~2 initially. In interaction picture, the systematic Hamiltonian under resonant driving is ($\hbar=1$)
\begin{eqnarray}\label{e1}
\hat H_I=\Big[\sum_{j=1,2}\Omega(t)|1\rangle_j\langle r|+{\rm H.c.}\Big]+\hat H_{rr},
\end{eqnarray}
where $\hat H_{rr}=V|rr\rangle_j\langle rr|$ is the RRI Hamiltonian, $\Omega(t)$ is the time-dependent Rabi frequency. Equation~(\ref{e1}) is traditionally regarded as a blockade Hamiltonian because $|rr\rangle$ state is blockaded when the two Rydberg atoms are driving simultaneously under the condition $V\gg {\rm max}\{\Omega(t)\}$. However, in this manuscript, we show that Hamiltonian~(\ref{e1}) with frequency-matched~(resonant) laser driving can break the blockade regime via modulating the time-dependent Rabi frequency. Another interesting thing is that the approximated form of Eq.~(\ref{e1}) reveals the concise Rabi oscillation between $|11\rangle$ and $|rr\rangle$ and thus could be used to construct the two-qubit quantum logic gate in one step without introducing other operations, such as extra laser driving~\cite{Su2016}.

\emph{The proposed RAB.---}
The required time-dependent Rabi frequency is given as a harmonic form
\begin{equation}\label{Omega}
\Omega(t)=\Omega_{\rm m}\cos(\omega t)
\end{equation}
with $\Omega_{\rm m}$ being the maximum amplitude and $\omega$ the oscillating angular frequency, which can be realized experimentally by acousto-optic modulator~(AOM)~\cite{dugan1997high,du2016experimental,diao2018shortcuts,PhysRevA.97.013628} or electro-optic modulator~(EOM)~\cite{weidner2018simplified} with the help of the arbitrary
waveform generator. In this case, through using Euler formula Hamiltonian~(\ref{e1}) can be rewritten as
\begin{eqnarray}\label{e2}
\hat H_I=\Big[\sum_{j=1,2}\frac{\Omega_{\rm m}}2(e^{i\omega t}+e^{-i\omega t})|1\rangle_j\langle r|+{\rm H.c.}\Big]+\hat H_{rr}.
\end{eqnarray}
In other words, the transition $|1\rangle\rightarrow|r\rangle$ driven resonantly by a single laser with time-dependent Rabi frequency in Eq.~(\ref{Omega}) is equivalent to that driven dispersively by two classical fields with the same Rabi frequency $\Omega_{\rm m}/2$ but with opposite detuning.
Besides, the blue-detuned dispersive interaction can be used to compensate RRI-(with positive value)-induced energy shift~\cite{Ates2007,Pohl2009,qian2009,Amthor2010,Zuo2010,Lee2012,Carr2013,Li2013,Su2016,*Su201702,Su2018,Shao2017,chen2018accelerated,Li2019} and further realize the RAB. 

For convenience, we investigate the dynamics of the two-atom system in the rotation frame with respect to $\hat H_{rr}$ under the two-atom basis $|mn\rangle$~($m,n=0,1,r$). 
Then, if the parameter condition $V=2\omega$ and $\omega\gg \Omega_{\rm m}/2$ are satisfied, the effective form of Hamiltonian~(\ref{e2}) could be achieved as
\begin{eqnarray}\label{e4}
\hat{H}_{\rm eff}=\Big(\frac{\Omega_{\rm m}^2}{2\omega}|11\rangle\langle rr|+ {\rm H.c.}\Big)+\frac{2\Omega_{\rm m}^2}{3\omega}|rr\rangle\langle rr|,
\end{eqnarray}
based on the second-order perturbation theory~\cite{James2007}, in which we have neglected the single excitation terms that are decoupled to the two-qubit space $\{|kl\rangle\}$~($k,l=0,1$). The effective Hamiltonian~(\ref{e4}) involves only the $|11\rangle\leftrightarrow|rr\rangle$ interaction, in spite of the Stark shift of $|rr\rangle$. Alternatively, the Stark shift of $|rr\rangle$ could be absorbed into the RRI by modifying the RRI strength as 
\begin{equation}\label{modified}
V=2\omega-{2\Omega_{\rm m}^2}/{3\omega}.
\end{equation}
Therefore, the resulting effective Hamiltonian~(\ref{e4}) becomes
\begin{eqnarray}\label{e5}
\hat{H}_{\rm e}=\frac{\Omega_{\rm m}^2}{2\omega}|11\rangle\langle rr|+ {\rm H.c.}.
\end{eqnarray}
When the initial state of the two-atom system is $|11\rangle$, the state at time \emph{t} can be obtained by solving the Schr\"{o}dinger equation $i\partial_t|\psi(t)\rangle=\hat{H}_{\rm e}|\psi(t)\rangle$ as
\begin{eqnarray}\label{e6}
|\psi(t)\rangle=\cos\Big(\frac{\Omega_{\rm m}^2t}{2\omega}\Big)|11\rangle-i\sin\Big(\frac{\Omega_{\rm m}^2t}{2\omega}\Big)|rr\rangle.
\end{eqnarray}
The two-atom simultaneous excitation can be gotten at the time $t=(2n-1)\pi\omega/\Omega_{\rm m}^2$ with $n$ being a positive integer.

\begin{figure}
\includegraphics[width=\linewidth]{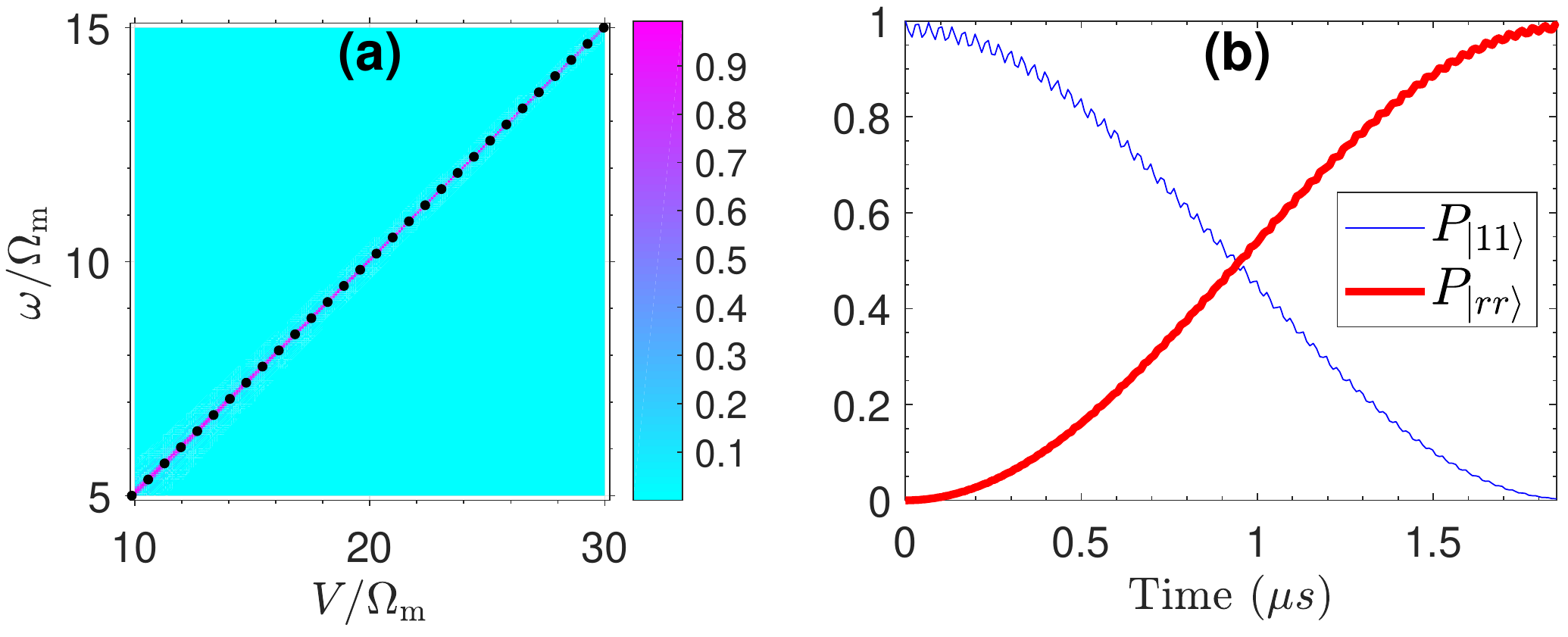}\\
\caption{(a)~Population of $|rr\rangle$ versus $V/\Omega_{\rm m}$ and $\omega/\Omega_{\rm m}$ at the time $t=\pi\omega/\Omega_{\rm m}^2$; Solid dots are plotted through the equation $V=2\omega-{2\Omega_{\rm m}^2}/{3\omega}$. 
(b)~Population inversion between $|11\rangle$ and $|rr\rangle$, with parameters $\Omega_{\rm m}/2\pi=2$~MHz, $\omega=7.5\Omega_{\rm m}$, and $V=2\omega-{2\Omega_{\rm m}^2}/{3\omega}$. To clearly see the validity of the RAB and the systematic dynamics, here we set $\gamma=0$ for simplicity.}\label{F2}
\end{figure}
In order to get the concise Hamiltonian in Eq.~(\ref{e5}), the parameter condition $V=2\omega-{2\Omega_{\rm m}^2}/{3\omega}$ is introduced. In Fig.~\ref{F2}(a), we simulate the population of $|rr\rangle$ versus $V/\Omega_{\rm m}$ and $\omega/\Omega_{\rm m}$ at the time $t=\pi\omega/\Omega_{\rm m}^2$ by solving numerically the following Lindblad master equation
\begin{eqnarray}\label{e7}
\dot{\hat{\rho}}(t)&=&i[\hat{\rho}(t)\hat{H}_I-\hat{H}_I\hat{\rho}(t)]+\frac{1}{2}\sum^{4}_{k=1}\{2\mathcal{\hat{L}}_{k}\hat{\rho}(t)\mathcal{\hat{L}}_{k}^{\dag}
\nonumber\\
&&-[\mathcal{\hat{L}}_{k}^{\dag}\mathcal{\hat{L}}_{k}\hat{\rho}(t)+\hat{\rho}(t)\mathcal{\hat{L}}_{k}^{\dag}\mathcal{\hat{L}}_{k}]\},
\end{eqnarray}
where $\mathcal{\hat{L}}_{1}\equiv\sqrt{\gamma/2}|0\rangle_{1}\langle r|$, $\mathcal{\hat{L}}_{2}\equiv\sqrt{\gamma/2}|1\rangle_{1}\langle r|$, $\mathcal{\hat{L}}_{3}\equiv\sqrt{\gamma/2}|0\rangle_{2}\langle r|$, and $\mathcal{\hat{L}}_{4}\equiv\sqrt{\gamma/2}|1\rangle_{2}\langle r|$ are Lindblad operators describing the four decay paths.
Here the population of the quantum state $|\varphi\rangle$ is defined as 
\begin{equation}\label{population}
P_{|\varphi\rangle}(t)\equiv\langle\varphi|\hat{\rho}(t)|\varphi\rangle, 
\end{equation}
in which $\hat{\rho}(t)$ denotes the density operator of the two-atom system at time \emph{t}. From Fig.~\ref{F2}(a), one can see that the line of the near-unity population of $|rr\rangle$~(numerical result) is well coincident with the line $V=2\omega-{2\Omega_{\rm m}^2}/{3\omega}$~(analytical result). In Fig.~\ref{F2}(b), we plotted the population inversion between $|11\rangle$ and $|rr\rangle$ with the system initially in $|11\rangle$. The result shows that the population of $|rr\rangle$ is very close to 1 at the time $t=\pi\omega/\Omega_{\rm m}^2$, which means the RAB is realized completely.

Up till now, we have analytically shown that the RAB could be achieved in one step via modulating the time-dependent Rabi frequency of the resonant interaction, and analyzed the RAB through numerically solving the master equation. In the following, we would show the applications of the proposed RAB on the construction of the quantum logic gate.

\emph{Applications in quantum logic gates.---} We now discuss the applications of the proposed unconventional RAB in quantum logic gate. Firstly we show how  the RAB could be used to construct the CZ gate. The form of CZ gate is 
\begin{equation}
\hat{U}_{\rm CZ}=|00\rangle\langle00|+|01\rangle\langle01|+|10\rangle\langle10|-|11\rangle\langle11|.
\end{equation}
From Hamiltonian~(\ref{e5}) and the evolution process in Eq.~(\ref{e6}), one can see that $|11\rangle\rightarrow-|11\rangle$ can be implemented at time $T=2(2n-1)\pi\omega/\Omega_{\rm m}^2$~(\emph{n} denotes positive integer) in one step while the other three states $|00\rangle$, $|01\rangle$, and $|10\rangle$ always keep invariant during the whole process because they are decoupled with the Hamiltonian~(\ref{e5}). That is, the CZ gate can be constructed in one step 

In addition to the CZ gate, the CNOT gate can also be constructed by introducing the other classical laser imposed on Atom~2 and resonantly driving the transition $|0\rangle_2\leftrightarrow|r\rangle_2$ with Rabi frequency $-\Omega(t)$. On that basis, the two-atom Hamiltonian in the interaction picture should be rewritten as
\begin{eqnarray}\label{e8}
\mathcal{\hat H_I}=\hat H_I-[\Omega(t)|0\rangle_2\langle r|+{\rm H.c.}],
\end{eqnarray}
in which $\hat H_I$ is shown in Eq.~(\ref{e1}). Similar to the derivation process from Eq.~(\ref{e1}) to Eq.~(\ref{e5}), an effective form of Hamiltonian $\mathcal{\hat H_I}$ can be obtained as
\begin{eqnarray}\label{e9}
\mathcal{\hat{H}_{\rm e}}=\frac{\Omega_{\rm m}^2}{2\omega}(|11\rangle-|10\rangle)\langle rr|+ {\rm H.c.}.
\end{eqnarray}
And the modified relation in Eq.~(\ref{modified}) should be changed to
\begin{equation}\label{modified1}
V=2\omega-{\Omega_{\rm m}^2}/{\omega}.
\end{equation}
Then by solving analytically the Schr\"{o}dinger equation with Hamiltonian $\mathcal{\hat{H}_{\rm e}}$, one can get the evolution as
\begin{eqnarray}\label{e10}
|11\rangle\rightarrow\cos^2\theta|11\rangle-\frac{i}{\sqrt2}\sin(2{\theta})|rr\rangle+\sin^2\theta|10\rangle,\nonumber\\
|10\rangle\rightarrow\cos^2\theta|10\rangle+\frac{i}{\sqrt2}\sin(2{\theta})|rr\rangle+\sin^2\theta|11\rangle,
\end{eqnarray}
with $\theta\equiv{\Omega_{\rm m}^2t}/{2\sqrt2\omega}$. By choosing the operation time $T=\sqrt2(2n-1)\pi\omega/\Omega_{\rm m}^2$ with \emph{n} being positive integer, the CNOT gate
\begin{eqnarray*}
\hat{U}_{\rm CNOT}=|00\rangle\langle00|+|01\rangle\langle01|+|10\rangle\langle11|+|11\rangle\langle10|
\end{eqnarray*}
would be constructed successfully in one step.

\begin{figure}
\includegraphics[width=\linewidth]{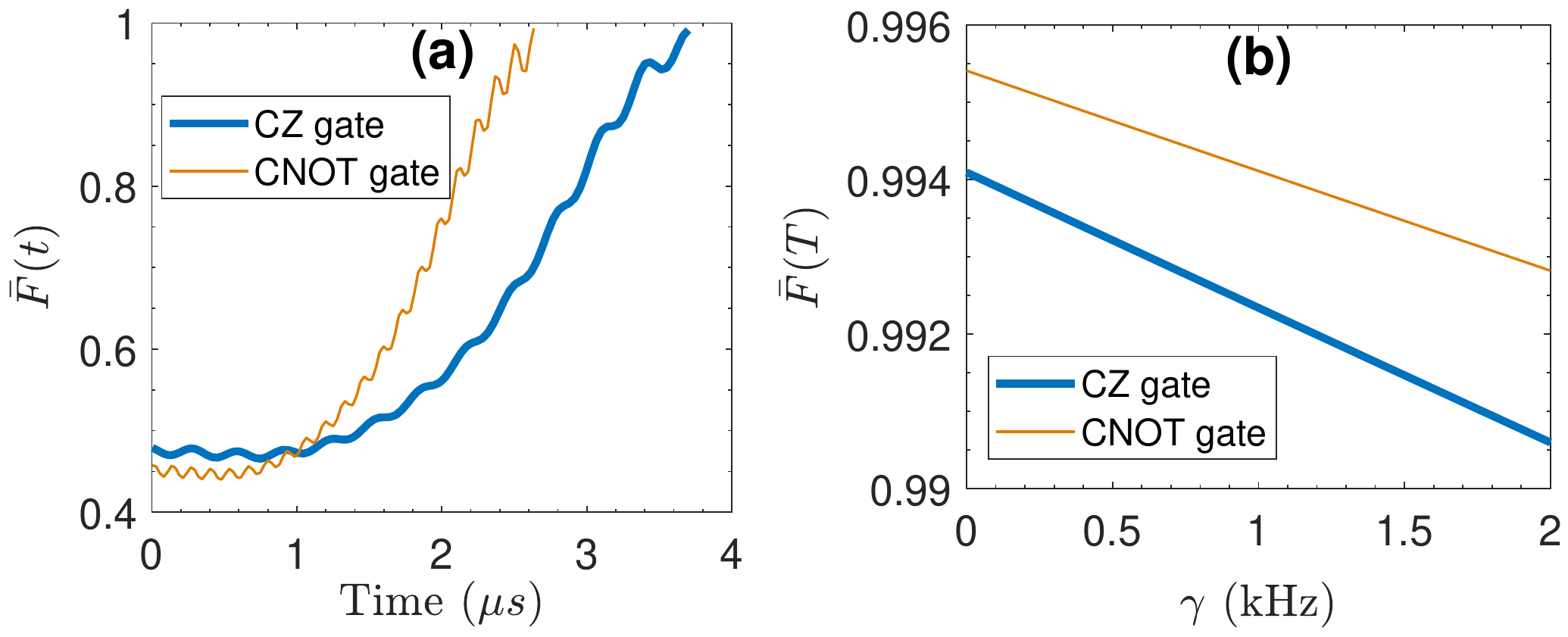}\\
\caption{(a)~Time-dependence of the average fidelity of the CZ gate~(thick line) and the CNOT gate~(thin line) with $\gamma=1.5$~KHz;
(b)~Influence of different atomic decay rates on the final average fidelity of the CZ gate~(thick line) and the CNOT gate~(thin line). Parameters used here: $\Omega_{\rm m}/2\pi=2$~MHz, $\omega=7.5\Omega_{\rm m}$, $\{V=2\omega-{2\Omega_{\rm m}^2}/{3\omega}$, $T=2\pi\omega/\Omega_{\rm m}^2\}$ for the CZ gate, and $\{V=2\omega-{\Omega_{\rm m}^2}/{\omega}$, $T=\sqrt2\pi\omega/\Omega_{\rm m}^2\}$ for the CNOT gate.}\label{F3}
\end{figure}
Through solving the master equation Eq.~(\ref{e7}) numerically~(as for constructing the CNOT gate, $\hat{H}_I$ should be replaced by $\mathcal{\hat H_I}$), we simulate the effectiveness of constructing the CZ gate and the CNOT gate, respectively, by means of the average gate fidelity, as shown in Fig.~\ref{F3}. Here we assume the two-atom system is initially in the direct-product state $|\Psi_i\rangle=(\cos\alpha|0\rangle_1+\sin\alpha|1\rangle_1)\otimes(\cos\beta|0\rangle_2+\sin\beta|1\rangle_2)$.
The average fidelity of the quantum gate $\hat{U}$ is defined as
\begin{eqnarray}
\bar{F}(t)\equiv\frac{1}{(2\pi)^2}\int_0^{2\pi}\int_0^{2\pi}\langle\Psi_i|\hat{U}^\dag\hat{\rho}(t)\hat{U}|\Psi_i\rangle {\rm d}\alpha {\rm d}\beta.
\end{eqnarray}
Throughout the process of constructing the CZ~(or CNOT) gate, the two-atom Rydberg excited state $|rr\rangle$ acts a vital intermediate role for the phase or qubit flip of the computational state though the Rydberg state $|r\rangle$ is not used to encode qubit, which exhibits the significance of the RAB regime for constructing quantum logical gates. Because the double-excitation state $|rr\rangle$ would be populated during constructing the logical gates, the atomic decay from $|r\rangle$ to a ground state $|0\rangle$ or $|1\rangle$ is supposed to be taken into account to assess the performance of the logical operation.

In Fig.~\ref{F3}(a), we consider the spontaneous emission rate $\gamma=1.5$~KHz to simulate the time-dependent evolution of the average fidelity of the CZ and the CNOT gate, respectively. With the time evolution, the average fidelity of the CZ~(CNOT) gate rises gradually and reaches up to near unity $\bar F=0.9915$~($\bar F=0.9935$) at the final time $T=2\pi\omega/\Omega_{\rm m}^2$~($T=\sqrt2\pi\omega/\Omega_{\rm m}^2$). The acceptable slight fluctuations are from the high-frequency oscillation errors and the average effect of gate fidelity. After all, $|rr\rangle$ is out of the computation space of the two-qubit logical gates and not populated in the initial or final state at all, so the final average gate fidelity can be over 0.99 with $\gamma=1.5$~KHz, which indicates that the protocol is robust against the atomic spontaneous emission. In order to further investigate the destructive effect of the atomic decay, we simulate the influence of different atomic decay rates on the final average fidelity of the CZ gate and the CNOT gate in Fig.~\ref{F3}(b). We clearly see that the atomic decay rate possesses a linear destruction effect on the final average fidelity. However, the final average fidelity decreases just by less than 0.004 when the atomic decay rate increases from 0 to $\gamma=2$~KHz. Even when $\gamma=2$~KHz$\sim10^{-3}\Omega_{\rm m}$, the final average fidelity of the CZ gate or the CNOT gate is still over 0.99. The atomic decay rate is relatively small due to the long-lived property of Rydberg state~\cite{Saffman2010}, and it is not an intractable challenge to keep the parameter relation $\gamma/\Omega_{\rm m}<10^{-3}$~\cite{Muller2014} that ensures an over-0.99 final average gate fidelity.

In conclusion, based on the single resonant classical field with time-dependent Rabi frequency, a one-step resonant-interaction-induced RAB has been proposed and the applications in quantum logic gate has also been studied. In shark contrast to the conventional two types of RAB regimes, neither the strong driving condition~($\Omega\gg V$)~\cite{PhysRevLett.85.2208} nor the dispersive interaction~\cite{Zuo2010,Lee2012,Carr2013,Li2013,Su2016,*Su201702,Su2018,Shao2017,chen2018accelerated,Li:18,Li2019} is required for the proposed unconventional RAB. In contrast to the conventional-RAB-based schemes that require introducing another laser driving to cancel the stark shifts~\cite{Su2018,Su2016} of the ground state or require multiple steps~\cite{Su201702}, the quantum logical gates based on the proposed RAB is more simple because no extra Stark shift is induced by the resonant driving. Numerical simulations show that high-fidelity CZ gate and CNOT gate can be constructed in one step under certain conditions. In addition, the complexity for constructing quantum logical gates is reduced because less external fields and controls are required.

\section*{ACKNOWLEDGEMENTS}
This work was supported by National Natural Science Foundation of China under Grants Nos. 11804308 and 11675046, and China Postdoctoral Science Foundation under No. 2018T110735.

\bibliographystyle{apsrev4-1}
\bibliography{refs}

\end{document}